# Modeling operational risk data reported above a time-varying threshold

**Pavel V. Shevchenko**
CSIRO Mathematical and Information Sciences, Sydney, Locked bag 17, North Ryde, NSW, 1670, Australia. e-mail: Pavel.Shevchenko@csiro.au

**Grigory Temnov** *
Edgeworth Centre for Financial Mathematics, School of Mathematical Sciences, University College Cork, Ireland, e-mail: g.temnov@ucc.ie

**Abstract**
Typically, operational risk losses are reported above a threshold. Fitting data reported above a constant threshold is a well known and studied problem. However, in practice, the losses are scaled for business and other factors before the fitting and thus the threshold is varying across the scaled data sample. A reporting level may also change when a bank changes its reporting policy. We present both the maximum likelihood and Bayesian Markov chain Monte Carlo approaches to fitting the frequency and severity loss distributions using data in the case of a time varying threshold. Estimation of the annual loss distribution accounting for parameter uncertainty is also presented.

\* A significant part of the work was made within "PRisMa Lab" in Vienna University of Technology
   Final parts of the paper, from the side of the second author, were completed at the Edgeworth Centre
    for Financial Mathematics, SFI research grant 07/MI/008, in University College Cork

# 1. Introduction

The Basel II Accord requires banks to meet a capital requirement for operational risk as part of an overall risk-based capital framework; see BIS (2006). To estimate the capital charge, many banks adopt the Loss Distribution Approach (LDA) under the Basel II Advanced Measurement Approaches (AMA). The LDA is based on estimation of the severity and frequency distributions of the loss events for each risk cell in a bank over a one year time horizon. The industry usually refers these risk cells as "risk nodes" in the Basel II regulatory matrix of eight business lines by seven risk types. The capital charge for operational risk is then based on the 0.999 quantile of the annual loss distribution.

Accurate modelling of the severity and frequency distributions is the key to estimating a capital charge. There are various important aspects of operational risk modeling discussed in the literature, e.g. Chavez-Demoulin *et al.* (2006), Cruz (2002, 2004) and Shevchenko (2009) to mention a few. One of the challenges in modelling operational risk is the lack of complete data – often a bank's internal data are not reported below a certain level (typically of the order of €10,000). These data are said to be left-truncated. Generally speaking, missing data increase uncertainty in modelling. Sometimes a threshold level is introduced to avoid difficulties with collection of too many small losses. Industry data are available through external databases from vendors (e.g. Algo OpData provides publicly reported operational risk losses above US$1million) and consortia of banks (e.g. ORX provides operational risk losses above €20,000 reported by ORX members). Several Loss Data Collection Exercises (LDCE) for historical operational risk losses over many institutions were conducted and their analyses reported in the literature. In this respect, two papers are of high importance: Moscadelli (2004) analysing 2002 LDCE and Dutta and Perry (2006) analysing 2004 LDCE where the data were mainly above €10,000 and US$10,000 respectively.

Often, modelling of missing data is done assuming a parametric distribution for losses below and above the threshold. Then fitting is accomplished using losses reported above the threshold via the maximum likelihood method (see e.g. Frachot, Moudoulaud and Roncalli (2004)) or the Expectation Maximization algorithm (see e.g. Bee (2005)). The effect of data truncation in operational risk was studied in Baud, Frachot and Roncalli (2003), Chernobai *et al.* (2005), Mignola and Ugoccioni (2006), and Luo *et al.* (2007).

Typically the case of a constant threshold is discussed in research studies. In this paper we consider the case of a threshold level varying across observations. One of the reasons for varying threshold in operational risk loss data is that the losses are scaled for inflation and other factors before fitting, to reflect changes in risk over time. The reporting level may also change from time to time within a bank when reporting policy is changed. The problem with multiple thresholds also appears when different companies report losses into the same database using different threshold levels; see Baud *et al*. (2002).

One common practice in calculating the annual loss distribution is to ignore the uncertainty in the fitted parameters. That is, the distribution conditional on the fitted parameters is used to estimate quantiles and final capital charge. Ignoring this uncertainty, which is always present in loss data modelling, may lead to a significant underestimation of capital charge. The uncertainty of parameter estimates can be treated efficiently using a Bayesian framework; see Shevchenko and Wüthrich (2006), and Shevchenko (2008) for application of Bayesian inference in the operational risk context.

In this paper we use the Bayesian Markov chain Monte Carlo (MCMC) procedure that estimates not only parameter point estimators but also the distribution of the parameter errors. This allows for quantification of the posterior distribution for the parameters and annual loss distribution accounting for the process and parameter uncertainties. Typically, practitioners in operational risk regard MCMC methods as difficult to implement and use. Here, for illustrative purpose, we demonstrate the use of a Random Walk Metropolis Hastings (RWMH) within Gibbs algorithm which is very efficient for estimation of the distribution parameters when the likelihood can be calculated efficiently. If the likelihood is difficult to evaluate as in the cases of modeling many risks that follow Poisson processes with dependent intensities (see Peters *et al*. (2009b)) or severity densities that do not have closed form (e.g. g-and-h distribution), more advanced MCMC methods such as Slice Sampler or Approximate Bayesian Computation can be used; see Peters and Sisson (2006) in the operational risk context. Also, in general, questions of Bayesian model choice must be addressed; see Peters and Sisson (2006).

We do not consider real data. The main objective of the paper is to demonstrate how to model data reported above a time varying threshold and illustrate the use of the RWMH within Gibbs algorithm.

The organisation of this paper is as follows. Section 2 describes the models of constant and varying in time thresholds. Section 3 and Section 4 describe Bayesian framework and MCMC procedures. Section 5 presents numerical results. Discussions and conclusions are presented in Section 6.

## 2. Model

Hereafter, we consider a single risk cell only. To clarify notation, we shall use upper case symbols to represent random variables, lower case symbols for their realizations and bold for vectors. The commonly used LDA for operational risk assumes that loss events follow some point process so that the annual loss in a risk cell in year *m* is

$$Z_m = \sum_{i=1}^{N_m} X_i(m). \tag{1}$$

Here, $N_m$ is the number of events (frequency) and $X_i(m)$, $i = 1,...,N_m$ are the severities of the events in year *m*. Typically it is assumed that the loss events are modelled by a homogeneous Poisson process with the intensity parameter $\lambda$. Thus, $N_m$, $m = 1,2,...$ are independent and identically distributed (iid) random variables (rvs) from the Poisson distribution, $Poisson(\lambda)$, with

$$\Pr[N_m = n] = p(n \mid \lambda) = \frac{\lambda^n}{n!} \exp[-\lambda], \quad \lambda > 0, n = 0,1,... \tag{2}$$

and the event inter-arrival times $\delta T_j = T_j - T_{j-1}$, $j = 1,2,...$ (where $T_0 < T_1 < T_2 < ...$ are the event times and $T_0 = t_0$ is the start of the observation period) are iid rvs from an exponential distribution with the pdf and cdf

$$g(\tau \mid \lambda) = \lambda \exp(-\lambda \tau) \text{ and } G(\tau \mid \lambda) = 1 - \exp(-\lambda \tau) \tag{3}$$

respectively. General properties of Poisson processes are discussed in many textbooks, see e.g. Daley and Vere-Jones (1988). The severities of the events $X_j$, $j = 1,2,...$ which occur at $T_j$, $j = 1,2,...$ respectively, are modelled as iid rvs from a continuous distribution $F(x \mid \boldsymbol{\beta})$, $0 < x < \infty$, whose density is denoted as $f(x \mid \boldsymbol{\beta})$. Here, $\boldsymbol{\beta}$ are the severity distribution parameters. Hereafter, it is assumed that the severities and frequencies of the events are independent. If convenient, we may index severities $X_j$ and event times $T_j$, $j = 1, 2, \ldots$ as $X_i(m)$ and $T_i(m)$, $i = 1, \ldots, N_m$, $m = 1, 2, \ldots$ respectively.

It is implicitly assumed that considered data for the severities and frequencies correspond to the losses before insurance policy (if any) is applied. If details of the insurance policies are known (e.g. top cover limit, excess amount, etc), see Shevchenko (2009), then it should not be difficult to account for insurance recoveries, when the annual loss distribution over next reporting year is estimated via simulation of the frequencies and severities. Often, event times are needed to calculate the insurance recoveries and these are easily simulated for Poisson processes considered in this paper.

## *2.1. Constant threshold: likelihood and MLEs*

If the losses originating from $f(x|\boldsymbol{\beta})$ and $p(k|\lambda)$ are recorded above a known reporting level (truncation level) $L$, then the density of the losses above $L$ is left-truncated

$$f_L(x|\boldsymbol{\beta}) = \frac{f(x|\boldsymbol{\beta})}{1 - F(L|\boldsymbol{\beta})}; \quad L \leq x < \infty. \tag{4}$$

The events of the losses above $L$ follow a Poisson process with the intensity

$$\theta(\boldsymbol{\gamma}, L) = \lambda \times (1 - F(L|\boldsymbol{\beta})), \tag{5}$$

a so-called thinned Poisson process, and the annual number of events above the threshold is distributed as $Poisson(\theta)$. Hereafter, $\boldsymbol{\gamma} = (\lambda, \boldsymbol{\beta})$ is a vector of all distribution parameters

Consider a random vector $\mathbf{Y}$ of the events recorded above the threshold $L$ over a period of $M$ years consisting of the annual frequencies $\widetilde{N}_m$, $m = 1, \ldots, M$ and severities $\widetilde{X}_j$, $j = 1, \ldots, J$, $J = \widetilde{N}_1 + \ldots + \widetilde{N}_M$. The joint density (likelihood) of $\mathbf{Y}$ at $\widetilde{N}_m = \widetilde{n}_m$ and $\widetilde{X}_j = \widetilde{x}_j$ can be written as

$$\ell(\mathbf{y}|\boldsymbol{\gamma}) = \prod_{j=1}^{J} f_L(\widetilde{x}_j|\boldsymbol{\beta}) \prod_{m=1}^{M} p(\widetilde{n}_m|\theta(\boldsymbol{\gamma}, L)). \tag{6}$$

Then the maximum likelihood estimators (MLEs) $\hat{\boldsymbol{\gamma}}$ can be found as a solution of

$$\begin{aligned}
\frac{\partial \ln \ell(\mathbf{y}|\boldsymbol{\gamma})}{\partial \lambda} &= [1 - F(L|\boldsymbol{\beta})] \sum_{m=1}^{M} \frac{\partial}{\partial \theta} \ln p(\widetilde{n}_m|\theta(\boldsymbol{\gamma}, L)) = 0, \\
\frac{\partial \ln \ell(\mathbf{y}|\boldsymbol{\gamma})}{\partial \boldsymbol{\beta}} &= \sum_{j=1}^{J} \frac{\partial}{\partial \boldsymbol{\beta}} \ln f_L(\widetilde{x}_j|\boldsymbol{\beta}) - \lambda \frac{\partial F(L|\boldsymbol{\beta})]}{\partial \boldsymbol{\beta}} \sum_{m=1}^{M} \frac{\partial}{\partial \theta} \ln p(\widetilde{n}_m|\theta(\boldsymbol{\gamma}, L)) = 0.
\end{aligned} \tag{7}$$

It is easy to see that the MLEs $\hat{\boldsymbol{\beta}}$ for the severity parameters can be found marginally (independently from frequency) by maximizing

$$\sum_{j=1}^{J} \ln f_L(\widetilde{x}_j|\boldsymbol{\beta}) \tag{8}$$

and then the first equation in (7) gives the MLE for the intensity

$$\hat{\lambda} = \frac{1}{[1 - F(L | \hat{\boldsymbol{\beta}})]} \frac{1}{M} \sum_{m=1}^{M} \tilde{n}_m. \tag{9}$$

Similarly, if the data $\mathbf{Y}$ of the events above a constant threshold over the time period $[t_0, t_E]$ consist of the event inter-arrival times $\delta \tilde{T}_j = \tilde{T}_j - \tilde{T}_{j-1}$, $j = 1, ..., J$ (where $\tilde{T}_j$, $j = 1, 2, ..$ are the event times and $\tilde{T}_0 = t_0$) and the severities $\tilde{X}_j$, $j = 1, ..., J$, then the joint density (likelihood) of $\mathbf{Y}$ at $\delta \tilde{T}_j = \tilde{\tau}_j$ and $\tilde{X}_j = \tilde{x}_j$

$$\ell(\mathbf{y} | \boldsymbol{\gamma}) = [1 - G(t_E - \tilde{\tau}_J | \theta(\boldsymbol{\gamma}, L))] \prod_{j=1}^{J} f_L(\tilde{x}_j | \boldsymbol{\beta}) \times g(\tilde{\tau}_j | \theta(\boldsymbol{\gamma}, L))$$

$$= \lambda^J \exp(-\theta(\boldsymbol{\gamma}, L)(t_E - t_0)) \prod_{j=1}^{J} f(\tilde{x}_j | \boldsymbol{\beta}), \tag{10}$$

where $[1 - G(t_E - \tilde{\tau}_J | \theta(\boldsymbol{\gamma}, L))]$ is the probability that no event will occur within $(\tilde{\tau}_J, t_E]$. Then it is easy to see that the MLEs $\hat{\boldsymbol{\beta}}$ for the severity parameters are obtained by maximizing $\sum_{j=1}^{J} \ln f_L(\tilde{x}_j | \boldsymbol{\beta})$ with respect to $\boldsymbol{\beta}$ and the intensity MLE is

$$\hat{\lambda} = \frac{J}{[1 - F(L | \hat{\boldsymbol{\beta}})](t_E - t_0)}, \tag{11}$$

which is equivalent to (9) if the start and end of the observation period correspond to the beginning and end of the first and last years respectively.

## 2.2. Threshold varying in time: likelihood and MLEs

Often, in practice, before fitting a specific severity distribution, a modeller scales the losses by some factors (inflation, business factors, etc). The reporting threshold should be scaled correspondingly and thus the losses in the fitted sample will have different threshold levels. To model this situation consider the following set up.

- In the absence of threshold, the events follow a homogeneous Poisson process with the intensity $\lambda$ and the severities $X_j$ are iid from $F(.|\boldsymbol{\beta})$.
- The losses are reported above the known time dependent level $L(t)$. Denote the severities and arrival times of the reported losses as $\tilde{X}_j$ and $\tilde{T}_j$, $j = 1, ..., J$ respectively and $t_0$ is the start of the observation period.

Under the above assumptions, the events above $L(t)$ follow a non-homogeneous Poisson process with the intensity

$$\theta(\boldsymbol{\gamma}, L(t)) = \lambda \times (1 - F(L(t) | \boldsymbol{\beta})). \tag{12}$$

Denote

$$\Lambda(t,h) = \int_{t}^{t+h} \theta(\gamma, L(x))dx. \tag{13}$$

Then, given that $(j\text{-}1)$th event occurred at $\tilde{t}_{j-1}$, the inter-arrival time for the $j$-th event $\delta\tilde{T}_j = \tilde{T}_j - \tilde{T}_{j-1}$ is distributed from

$$G_j(\tau | \gamma) = 1 - \exp(-\Lambda(t_{j-1}, \tau)) \tag{14}$$

with the density

$$g_j(\tau | \gamma) = \theta(\gamma, L(t_{j-1} + \tau))\exp(-\Lambda(t_{j-1}, \tau)). \tag{15}$$

The number of events in year $m$ is $Poisson(\Lambda(s_m, 1))$ distributed, where $s_m$ is the time of the beginning of year $m$. Also note that the number of events from a non-homogeneous Poisson process over non-overlapping periods are independent.

The joint likelihood of the data $\mathbf{Y}$ of the events above $L(t)$ over the time period $[t_0, t_E]$, consisting of the inter-arrival times $\delta\tilde{T}_j = \tilde{T}_j - \tilde{T}_{j-1}$ and severities $\tilde{X}_j$, $j = 1,...,J$ above $L(t)$ can be written as

$$\ell(\mathbf{y}|\gamma) = [1 - G_J(t_E - \tilde{t}_J | \gamma)]\prod_{j=1}^{J} f_{L(\tilde{t}_j)}(\tilde{x}_j | \boldsymbol{\beta})g_j(\tilde{\tau}_j | \gamma). \tag{16}$$

Using (4) and (15), it is simplified to

$$\ell(\mathbf{y}|\gamma) = \lambda^J \exp(-\Lambda(t_0, t_E - t_0))\prod_{j=1}^{J} f(\tilde{x}_j | \boldsymbol{\beta}).$$

Here, explicitly, $\Lambda(t_0, t_E - t_0) = \lambda \int_{t_0}^{t_E}[1 - F(L(x)|\boldsymbol{\beta})]dx$. Then, the maximum likelihood equations are

$$\frac{\partial \ln \ell(\mathbf{y}|\gamma)}{\partial \lambda} = \frac{J}{\lambda} - \int_{t_0}^{t_E}[1 - F(L(x)|\boldsymbol{\beta})]dx = 0,$$

$$\frac{\partial \ln \ell(\mathbf{y}|\gamma)}{\partial \boldsymbol{\beta}} = -\frac{\partial}{\partial \boldsymbol{\beta}}\Lambda(t_0, t_E - t_0) + \sum_{j=1}^{J}\frac{\partial}{\partial \boldsymbol{\beta}}\ln f(\tilde{x}_j | \boldsymbol{\beta}) = 0. \tag{17}$$

The first equation gives

$$\hat{\lambda} = \frac{J}{\int_{t_0}^{t_E}[1 - F(L(x)|\hat{\boldsymbol{\beta}})]dx}. \tag{18}$$

This can be substituted into (16) and maximization will be required in respect to $\boldsymbol{\beta}$ only. The likelihood contains an integral over the severity distribution. If integration is not possible in closed form then it can be calculated numerically (that can be done efficiently using standard routines available in many numerical packages). For convenience, one can assume that a threshold is constant between the reported events: $L(t) = L(t_j)$, $\tilde{t}_{j-1} < t \leq \tilde{t}_j$ and $L(t) = L(t_E)$ for $\tilde{t}_j < t \leq t_E$, so that

$$\int_{t_0}^{t_E}[1 - F(L(x)|\boldsymbol{\beta})]dx = [1 - F(L(t_E)|\boldsymbol{\beta})](t_E - \tilde{t}_J) + \sum_{j=1}^{J}[1 - F(L(\tilde{t}_j)|\boldsymbol{\beta})]\tau_j . \qquad (19)$$

Of course this assumption is reasonable if the intensity of the events is not small. Typically scaling is done on the annual basis and one can assume piecewise constant threshold per annum and the integral is replaced by a simple summation.

The MLEs for severity parameters calculated marginally, i.e. by simply maximizing $\sum \ln f_{L(t_j)}(\tilde{x}_j|\boldsymbol{\beta})$, do not differ materially from the results of the joint estimation if the variability of the threshold is not extremely fast. The results of the estimation for the simulated data in the case of exponentially varying threshold, presented in section 5, confirm that intuitive observation, although the difference can still be significant if the intensity is small. Also, marginal estimation does not allow for quantification of the covariances between frequency and severity parameters required to account for parameter uncertainty.

If a data vector $\mathbf{Y}$ of the events above the reporting threshold consists of the annual counts $\tilde{N}_m$, $m = 1,...,M$ and severities $\tilde{X}_j$, $j = 1,...,J$ ($J = \tilde{N}_1 + ... + \tilde{N}_M$) then the joint likelihood of observations is

$$\ell(\mathbf{y}|\boldsymbol{\gamma}) = \prod_{j=1}^{J} f_{L(t_j)}(\tilde{x}_j|\boldsymbol{\beta}) \prod_{m=1}^{M} p(\tilde{n}_m|\Lambda(s_m,1)). \qquad (20)$$

Usually, in practice, scaling is done on an annual basis. Thus we can consider the case of piecewise constant threshold per annum such that for year $m$: $L(t) = L_m$, $\theta_m = \theta(\boldsymbol{\gamma}, L(t)) = \lambda(1 - F(L_m|\boldsymbol{\beta}))$, $s_m \leq t < s_m + 1$, where $s_m$ is the time of the beginning of year $m$. The likelihood in this case

$$\ell(\mathbf{y}|\boldsymbol{\gamma}) = \prod_{j=1}^{J} f_{L(\tilde{t}_j)}(\tilde{x}_j|\boldsymbol{\beta}) \prod_{m=1}^{M} p(\tilde{n}_m|\theta_m)) \qquad (21)$$

and equations to find MLEs are

$$\frac{\partial \ln \ell(\mathbf{y}|\boldsymbol{\gamma})}{\partial \lambda} = \sum_{m=1}^{M}[1 - F(L_m|\boldsymbol{\beta})]\frac{\partial}{\partial \theta_m}\ln p(\tilde{n}_m|\theta_m) = 0,$$

$$\frac{\partial \ln \ell(\mathbf{y}|\boldsymbol{\gamma})}{\partial \boldsymbol{\beta}} = \sum_{j=1}^{J}\frac{\partial}{\partial \boldsymbol{\beta}}\ln f_{L(\tilde{t}_j)}(\tilde{x}_j|\boldsymbol{\beta}) - \lambda\sum_{m=1}^{M}\frac{\partial F(L_m|\boldsymbol{\beta})}{\partial \boldsymbol{\beta}}\frac{\partial}{\partial \theta_m}\ln p(\tilde{n}_m|\theta_m) = 0. \qquad (22)$$

The first equation gives

$$\hat{\lambda} = \frac{\sum_{m=1}^{M} \tilde{n}_m}{\sum_{m=1}^{M} [1 - F(L_m | \hat{\boldsymbol{\beta}})]}. \quad (23)$$

This can be substituted into the likelihood function (21), so maximization to obtain the MLEs is required in respect to severity parameters only.

The MLEs of the severity parameters should be estimated jointly with the intensity. However, given that the intensity MLE can be expressed in terms of the severity parameters MLEs via (18) or (23), the maximization of the likelihood can be done effectively in respect to severity parameters only by substituting (18) into (16), or respectively substituting (23) into (21).

### 2.3. Non-homogeneous Poisson

The results of Section 2.2 can easily be extended to the case when the event process before truncation is a non-homogeneous Poisson process with the time dependent intensity $\lambda(t)$. In this case, after truncation, the events above $L(t)$ follow a non-homogeneous Poisson with the intensity

$$\lambda(t)(1 - F(L(t) | \boldsymbol{\beta})).$$

One has to simply change $\lambda$ to $\lambda(t)$ in (12-15), then the expressions for the likelihoods (16) and (20) are still valid. A parametric form can be assumed for $\lambda(t)$, for example:

$$\lambda(t) = \begin{cases} \lambda_1; & t < t_c, \\ \lambda_2; & t \geq t_c, \end{cases}$$

where $t_c$ could be a time of a new policy/control introduced; or one can consider the intensity as a function of some explanatory variables; etc. Then all unknown parameters can be fitted by the maximum likelihood or MCMC as described in Section 4. A particular parametric form is problem specific and our numerical examples below consider the case of constant intensity only.

If before the truncation, the process is a non-homogeneous Poisson $\lambda(t)$ and the truncation level is *unknown* function of time $L(t)$, then it might be a problem to distinguish the change in the intensity versus the change in the threshold. Though the importance of this case in practice is questionable. Certainly it depends on specific parametric models chosen for $\lambda(t)$ and $L(t)$ and will not be considered in this paper.

Significant for practical applications is the case when not just the threshold but also the losses themselves are subject to a certain trend with possibly unknown or misspecified parameters (so that the assumption of i.i.d. observations is violated). The likelihood of such a model would then include, besides the parameters of the severity and frequency, also the parameters associated with the trend. In that case, the optimization in the maximum likelihood method (or the MCMC modelling) should be made with respect to all the parameters of the model, including the trend parameters. Naturally, techniques for optimization and modelling regarding this type of model become significantly more complicated; consideration of this model is a complex problem and is beyond the scope of the present paper.

## *2.4. MLE asymptotic properties*

Often, as a sample size increases, the MLEs $\hat{\boldsymbol{\gamma}}$ have the following useful asymptotic properties:
- under the mild regularity conditions, $\hat{\boldsymbol{\gamma}}$ is a consistent estimator of the true parameter $\boldsymbol{\gamma}$, i.e. $\hat{\boldsymbol{\gamma}}$ converges to $\boldsymbol{\gamma}$ in probability;
- under the stronger regularity conditions, $\hat{\boldsymbol{\gamma}} - \boldsymbol{\gamma}$ has a Normal distribution with zero mean and covariance matrix $\mathbf{I}^{-1}(\boldsymbol{\gamma})$, where $I_{i,j}(\boldsymbol{\gamma}) = -E[\partial^2 \ln \ell(\mathbf{Y}|\boldsymbol{\gamma})/\partial \gamma_i \partial \gamma_j]$ is the Fisher information matrix.

If $\mathbf{I}(\boldsymbol{\gamma})$ can not be found in closed form, then (for a given realization $\mathbf{y}$) typically it is estimated by the observed information matrix

$$\hat{I}_{i,j}(\boldsymbol{\gamma}) = -\frac{\partial^2 \ln \ell(\mathbf{y}|\boldsymbol{\gamma})}{\partial \gamma_i \partial \gamma_j}.$$

Both $\hat{I}_{i,j}(\boldsymbol{\gamma})$ and $I_{i,j}(\boldsymbol{\gamma})$ depend on the unknown true parameter $\boldsymbol{\gamma}$, which is estimated by $\hat{\boldsymbol{\gamma}}$ in the final estimate of the covariances between MLEs

$$\mathrm{cov}(\hat{\gamma}_i, \hat{\gamma}_j) \approx \left(\hat{\mathbf{I}}^{-1}(\hat{\boldsymbol{\gamma}})\right)_{i,j}. \tag{24}$$

If closed form is not available then the required second order derivatives are calculated numerically using finite difference method. Precise regularity conditions required and proofs can be found in many textbooks; see e.g. Lehmann (1983) Theorem 6.4.1.

*Remarks*
- *The required regularity conditions for the above asymptotic theorem are conditions to ensure that the density is smooth with regard to parameters and there is nothing "unusual" about the density, see Lehmann (1983). These include that: the true parameter is an interior point of the parameter space; the density support does not depend on the parameters; the density differentiation with respect to the parameter and the integration over $\mathbf{y}$ can be swapped; third derivatives with respect to the parameters are bounded; and few others.*
- *Though the required conditions are mild, they are often difficult to be proved. Here, we just assume that these conditions are satisfied.*
- *Whether a sample size is large enough to use the asymptotic results is another difficult question that should be addressed in real applications.*

## 3. Bayesian Estimation

The parameters fitted using real data are estimates that have statistical fitting errors due to a finite sample size. The true parameters are not known. In our experience with banks, typically, uncertainty in fitted parameters is ignored when capital is quantified. That is parameters are fixed to their point estimates (e.g. maximum likelihood estimates) when the annual loss distribution and its 0.999 quantile are calculated. The Bayesian framework is convenient to account for parameter uncertainty on quantile estimates, see Shevchenko and Wüthrich (2006) for an application of the Bayesian framework to operational risk. Consider model (1) with a random vector of data (severities and frequencies) **Y** over *M* years. Given $\boldsymbol{\gamma} = (\lambda, \boldsymbol{\beta})$, denote the density of the annual loss as $h(z|\boldsymbol{\gamma})$. In the Bayesian approach the parameters $\boldsymbol{\gamma}$ are modelled as random variables. Then the density of the full predictive distribution for the next year annual loss $Z_{M+1}$ is

$$h(z|\mathbf{y}) = \int h(z|\boldsymbol{\gamma})\pi(\boldsymbol{\gamma}|\mathbf{y})d\boldsymbol{\gamma}, \quad (25)$$

where $\pi(\boldsymbol{\gamma}|\mathbf{y})$ is the joint posterior density of the parameters given data **Y**. From Bayes' rule

$$\pi(\boldsymbol{\gamma}|\mathbf{y}) \propto \ell(\mathbf{y}|\boldsymbol{\gamma})\pi(\boldsymbol{\gamma}), \quad (26)$$

where $\ell(\mathbf{y}|\boldsymbol{\gamma})$ is the likelihood of observations and $\pi(\boldsymbol{\gamma})$ is a prior distribution for the parameters (the prior distribution can be specified by an expert or fitted using external data or can be taken as uninformative so that inference is implied by data only).

The mode of the posterior distribution $\hat{\boldsymbol{\gamma}} = \text{mode}(\boldsymbol{\gamma})$ can be used as a point estimator for the parameters. For large sample size (and continuous prior distribution), it is common to approximate $\ln \pi(\boldsymbol{\gamma}|\mathbf{y})$ by a second-order Taylor series expansion around $\hat{\boldsymbol{\gamma}}$. Then $\pi(\boldsymbol{\gamma}|\mathbf{y})$ is approximately a multivariate Normal distribution with the mean $\hat{\boldsymbol{\gamma}}$ and covariance matrix calculated as the inverse of the matrix:

$$(\tilde{\mathbf{I}})_{i,j} = -\left.\frac{\partial^2 \ln \pi(\boldsymbol{\gamma}|\mathbf{y})}{\partial \gamma_i \partial \gamma_j}\right|_{\boldsymbol{\gamma}=\hat{\boldsymbol{\gamma}}}. \quad (27)$$

In the case of improper constant priors, i.e. $\pi(\boldsymbol{\gamma}|\mathbf{y}) \propto \ell(\mathbf{y}|\boldsymbol{\gamma})$, this approximation compares to the Gaussian approximation for the MLEs (24). Also, note that in the case of constant priors, the mode of the posterior and MLE are the same. This is also true if the prior is uniform within a bounded region, provided that the MLE is within this region.

Sometimes it is possible to find the posterior distribution $\pi(\boldsymbol{\gamma}|\mathbf{y})$ of the parameters in closed form. However, in general, $\pi(\boldsymbol{\gamma}|\mathbf{y})$ should be estimated numerically, e.g. using one of the Markov chain Monte Carlo methods.

### 3.1. The 0.999 quantile of the full predictive annual loss distribution

The 0.999 quantile $Q_{0.999}^B$ of the full predictive distribution $h(z|\mathbf{y})$ can be easily calculated, for example, by the following MC procedure:

**Step 1**. Simulate $\lambda$ and $\boldsymbol{\beta}$ from the joint distribution $\pi(\boldsymbol{\gamma}|\mathbf{y})$.

**Step 2**. Given $\lambda$ and $\boldsymbol{\beta}$, calculate the annual loss $z = \sum_{i=1}^{n} x_i$ by simulating $n$ from the frequency distribution $p(.|\lambda)$ and $x_i, i = 1,...,n$ from the loss severity distribution $F(.|\boldsymbol{\beta})$.

**Step 3**. Repeat **Step 1** to **Step 2** $K$ times to get realisations of total annual loss $z_j, j = 1,....,K$.

**Step 4**. Estimate the 0.999 quantile, $\hat{Q}_{0.999}^B$, of the total annual loss and its standard error (due to finite number of simulations) using simulated sample $z_j, j = 1,....,K$ in the usual way, see e.g. Shevchenko (2008).

In the above the total annual loss accounts for both the process uncertainty (severity and frequencies are random variables) and the parameter uncertainty (parameters are simulated from their posterior distribution). The parameter uncertainty comes from the fact that we do not know the true values of the parameters.

### 3.2. Distribution of the 0.999 quantile of the annual loss distribution

Another approach to account for parameter uncertainty is to consider a quantile $Q_{0.999}(\boldsymbol{\gamma})$ of the conditional annual loss distribution $h(z|\boldsymbol{\gamma})$. Then, given that $\boldsymbol{\gamma}$ is distributed from $\pi(\boldsymbol{\gamma}|\mathbf{y})$, one can find the distribution for $Q_{0.999}(\boldsymbol{\gamma})$ and form a one-sided or two-sided predictive interval to contain the true value of the quantile with some probability $q$. Then one can argue that the conservative estimate of the capital charge should be based on the upper bound of the constructed confidence interval. Calculation of the predictive interval for the quantile $Q_{0.999}(\boldsymbol{\gamma})$ of $h(z|\boldsymbol{\gamma})$ can be accomplished as follows:

**Step 1**. Simulate $\boldsymbol{\gamma} = (\lambda, \boldsymbol{\beta})$ from $\pi(\boldsymbol{\gamma}|\mathbf{y})$.

**Step 2**. Given $\lambda$ and $\boldsymbol{\beta}$ from the Step 1, calculate $Q_{0.999}(\boldsymbol{\gamma})$. Conceptually this can be accomplished via Monte Carlo: a) simulating $n$ from $p(.|\lambda)$ and $x_i, i = 1,...,n$ from $f(.|\boldsymbol{\beta})$; b) calculating the annual loss $z = \sum_{i=1}^{n} x_i$; c) repeating steps a) and b) many times to build a sample of annual losses used to estimate $Q_{0.999}(\boldsymbol{\gamma})$ in a usual way.

**Step 3**. Repeat Steps 1-2, $K$ times to build a sample of possible realizations of the quantile and use the sample to find the distribution of $Q_{0.999}(\boldsymbol{\gamma})$. Then a predictive interval to contain the true value of $Q_{0.999}$ with some probability $q$ can be constructed using the sample in the usual way.

The use of Monte Carlo in Step 2 to compute $Q_{0.999}(\boldsymbol{\gamma})$ for a given parameter vector $\boldsymbol{\gamma}$ is too computationally demanding and it is more efficient to use deterministic methods. One can choose well known Panjer recursion, Fast Fourier Transform or direct inversion of characteristic function methods. The analysis of these techniques and comparison with Monte Carlo method in the context of operational risk (see Temnov and Warnung (2008) and Luo and Shevchenko (2009)) shows that these methods are efficient for estimation of the 0.999 quantile, considering both the accuracy and the speed of calculation. In a recent paper, Peters, Johansen and Doucet (2007) proposed a hybrid Monte Carlo approach utilizing Panjer recursion, importance sampling and trans-dimensional Markov chain Monte Carlo.

    Note that this estimation requires a specification of a confidence level $q$, which could be difficult. One could choose $q = 0.95$ as a conservative estimate for the 0.999 quantile, as

most of statistical estimates use the 95% confidence level. However, in the context of operational risk the preferable confidence level should be in agreement with the requirements from regulator. Unless this level is set exactly, estimating the full predictive distribution for the annual loss and using its 0.999 quantile $Q^B_{0.999}$ for quantification of the capital charge as described in section 3.1 would be more appealing.

## 4. MCMC procedure

Simulation of parameters $\gamma = (\lambda, \boldsymbol{\beta})$ from the posterior distribution $\pi(\gamma | \mathbf{Y})$ required in procedures in Sections 3.1 and 3.2 can be accomplished using MCMC techniques. Here, we briefly outline Random Walk Metropolis Hastings (RWMH) within Gibbs scheme used for estimating the posterior distribution of parameters. In the context of operational risk, the RWMH within Gibbs is mentioned in Peters and Sisson (2006) and used in Peters *et al.* (2009a) for a similar problem in insurance. For other references concerning this algorithm, see Gelman *et al.* (1997), Bedard and Rosenthal (2008), Roberts and Rosenthal (2001), and Robert and Casella (2004).

The algorithm creates a reversible Markov chain with a stationary distribution corresponding to our target posterior distribution. It should be noted that the Gibbs sampler creates a Markov chain where each iteration involves scanning either deterministically or randomly over the variables that comprise the target stationary distribution of the chain. This process involves sampling each proposed parameter update from the corresponding full conditional posterior distribution.

In our calculations we assume that all parameters are independent under the prior density $\pi(\gamma)$ and distributed uniformly with $\gamma_i \sim U(a_i, b_i)$ on a wide ranges so that inference is mainly implied by data only (also see remark at the end of this section). Denote by $\gamma^{(k)}$ the state of the chain at iteration $k$ (usually the initial state $\gamma^{(k=0)}$ is taken as MLEs). Then the MCMC sampler proceeds by proposing to move the *i*th parameter from a state $\gamma_i^{(k-1)}$ to a new proposed state $\gamma_i^*$. The latter is sampled from a MCMC proposal transition kernel. As a proposal transition kernel, we use the Gaussian distribution truncated below $a_i$ and above $b_i$, with the density

$$f_N^{(T)}(\gamma_i^*; \gamma_i^{(k)}, \sigma_i) = \frac{f_N(\gamma_i^*; \gamma_i^{(k)}, \sigma_i)}{F_N(b_i; \gamma_i^{(k)}, \sigma_i) - F_N(a_i; \gamma_i^{(k)}, \sigma_i)}, \qquad (28)$$

where $f_N(x; \mu, \sigma)$ and $F_N(x; \mu, \sigma)$ are the Normal density and its distribution respectively with the mean $\mu$ and standard deviation $\sigma$. Then the proposed move is accepted with the probability

$$p(\gamma^{(k)}, \gamma^*) = \min\left\{1, \frac{\pi(\gamma^* | \mathbf{y}) f_N^{(T)}(\gamma_i^{(k-1)}; \gamma_i^*, \sigma_i)}{\pi(\gamma^{(k)} | \mathbf{y}) f_N^{(T)}(\gamma_i^*; \gamma_i^{(k-1)}, \sigma_i)}\right\}, \qquad (29)$$

where $\mathbf{y}$ is the vector of observations and $\pi(\gamma^* | \mathbf{y})$ is the posterior distribution. Also, here $\gamma^* = (\gamma_1^{(k)}, ..., \gamma_{i-1}^{(k)}, \gamma_i^*, \gamma_{i+1}^{(k-1)}, ...)$, i.e. $\gamma^*$ is a new state, where parameters $1, 2, ..., i-1$ are already updated while $i+1, i+2, ...$ are not updated yet.

The acceptance probability is a function of the transition kernel and the posterior density. Note that, a normalization constant for the posterior density is not needed here. If under the rejection rule one accepts the move then the new state of the *i*th parameter at iteration *k* is given by $\gamma_i^{(k)} = \gamma_i^*$, otherwise the parameter remains in the current state $\gamma_i^{(k)} = \gamma_i^{(k-1)}$ and an attempt to move that parameter is repeated at the next iteration.

In following this procedure, one builds a set of correlated samples from the target posterior distribution which have several asymptotic properties. One of the most useful of these properties is the convergence of ergodic averages constructed using the Markov chain samples to the averages obtained under the posterior distribution.

The chain has to be run until it has sufficiently converged to the stationary distribution (posterior distribution) and then one obtains samples from the posterior distribution. General properties of this algorithm, including convergence results can be found in e.g. Robert and Casella (2004) and application to a similar task in claims reserving is given in Peters *et al*. (2009a).

*Remarks*

- *The RWMH algorithm is simple in nature and easy to implement. However, if one does not choose the proposal distribution carefully, then the algorithm only gives a very slow convergence to the stationary distribution. There have been several studies regarding the optimal scaling of proposal distributions to ensure optimal convergence rates see e.g. Gelman et al. (1997), Bedard and Rosenthal (2007) and Roberts and Rosenthal (2001). According to the listed works, the asymptotic acceptance rate optimizing the efficiency of the process is 0.234 independent of the target density (for multivariate target distributions with i.i.d. components). In this case it is recommended that $\sigma_i$ are chosen to ensure that the acceptance probability is close to 0.234. To obtain this acceptance rate, one is required to perform some tuning of the proposal variance prior to final simulations.*
- *Recently, advanced modifications of MCMC algorithms were developed which allow on-line adaptations of optimal proposal distributions for specified mixing criteria. These algorithms are known as adaptive MCMC algorithms, and their use may significantly improve the effectiveness of the MCMC scheme. Adaptive MCMC algorithms can be used in our framework as well. For detailed descriptions and examples, we refer to Atchade and Rosenthal (2005) and Roberts and Rosenthal (2006).*
- *For practical use of the RWMH algorithm, it is important that the likelihood can be evaluated efficiently. This is usually the case if the severity density has a closed form (e.g. Pareto distribution considered in examples below, Lognormal distribution and many other standard distributions). In particular, it should not be a problem to use RWMH in the case of four-parameter GB2 distribution used in data analysis Dutta and Perry (2006). If the density can not be easily evaluated, while a simulation from the density is efficient (as in the case of another four-parameter distribution g-and-h used in Dutta and Perry (2006)), then more advanced MCMC methods such as Approximate Bayesian Computation can be used; see Peters and Sisson (2006).*

Summarizing this section, we list the most important quantities provided by MCMC and playing a role in the estimation of the aggregate loss quantiles:
- the mean of the posterior distribution (can be used as a point estimator for the parameters);
- the numerical standard error of the MCMC estimates (this quantity reflects the error in the point estimates of a parameter due to the finite number of chain iterations);
- the standard deviation of the posterior distribution (this uncertainty is associated with the finite size of the observed data sample).

Let us remark on a role of the choice of prior distributions in the Bayesian modelling. As we consider a basic model in the present work, we assume priors to be simply constant priors defined on some large region. This allows to get all inferences mainly implied by observations only. The constant priors used in this approach may be called *non-informative priors*, in contrast to a different approach associated with the choice of the so-called *informative priors*. In the latter approach, the choice of eligible informative priors plays an important role, as it has to be based on statistical properties of the data associated with priors. Informative priors can be used if external data and expert opinions are taken into account, see e.g. Lambrigger *et al* (2007). Strictly speaking, constant priors on a large region can be quite informative in some situations. In our numerical examples, we checked that the impact of chosen constant priors is not material by changing the region bounds.

## 5. Results

To illustrate the above procedures and calculations we perform a simulation experiment assuming a *Poisson*($\lambda$) process with Pareto distribution, *Pareto*($\alpha, \beta$), for the severities

$$F(x|\alpha,\beta) = 1 - (1 + x/\beta)^{-\alpha}, \quad x \geq 0, \alpha > 0, \beta > 0, \tag{30}$$

where $\alpha$ and $\beta$ are the shape and the scale parameters respectively. Then, for a reporting threshold varying in time $L(t)$, the distribution of the severity above the threshold occurred at time $t$ is a left-truncated Pareto distribution with pdf

$$f_{L(t)}(x|\alpha,\beta) = \frac{\alpha(1+x/\beta)^{-\alpha-1}}{\beta(1+L(t)/\beta)^{-\alpha}}; \quad L(t) \leq x < \infty. \tag{31}$$

The log-likelihood of the reported events (16) is

$$\ln \ell(\mathbf{y}|\boldsymbol{\gamma}) = J \ln(\alpha/\beta) - (1+\alpha)\sum_{j=1}^{J} \ln(1+x_j/\beta) + J \ln \lambda - \lambda \int_{t_0}^{t_E} (1+L(t)/\beta)^{-\alpha} dt. \tag{32}$$

In some special cases the integral $\lambda \int_{t_0}^{t_E} (1+L(t)/\beta)^{-\alpha} dt$ in (32) can be calculated explicitly but in general it will be calculated numerically. The latter might decrease the overall speed of the whole MCMC procedure as the integral has to be recalculated for each MCMC iteration. If the threshold is piecewise constant function of time (as assumed in our experiments below) then the integral is replaced by a simple summation. Also, in cases when the intensity of events is high enough, one can assume that the threshold is constant between events.

### 5.1. Simulation set up

We consider the observation period $0 \leq t \leq M$. Assume that the change of the threshold in time follows the exponential law, as this choice of the law for the varying threshold relies on the assumption that one of the most significant scaling factors for the losses is the inflation. If the real inflation rate is known, all the losses are scaled with respect to it, but this scaling cannot recover the missing data, which fell below the threshold and were not reported. Moreover, we assume that the scaling of the data is done per annum, therefore the value of the threshold stays constant each year $L(t) = L_m = L_0 \exp(r \times m)$, $m-1 \leq t < m$, $m = 1, 2, ...$ . Then, we simulate the loss events using the following procedure:

**Step 1**. Simulate the Poisson process event times $t_j$, $j = 1, 2...$ covering the period of $M$ years by simulating iid inter-arrival times $\tau_j = t_j - t_{j-1}$ from the exponential distribution with the parameter $\lambda$. Find the number of events $n_m$ occurred during each year $m = 1, ..., M$ and total number of events $n = n_1 + ... + n_M$.

**Step 2**. Simulate iid severities $x_j$ for the event times $t_j$, $j = 1, ..., n$ respectively from $Pareto(\alpha, \beta)$.

**Step 3**. Remove the events from the simulated sample when the event loss is below a varying threshold $L_m$.

**Step 4**. Given truncated sample of severities and event times, estimate the parameters $(\alpha, \beta, \lambda)$ via the MLE and MCMC procedures using likelihood (32).

**True parameter values.** We consider various observation periods $M$ and use the following parameters (unless indicated otherwise)
- Inflation rate $r = 0.03$;
- Pareto distribution shape and scale parameters $\alpha = 2.0$ and $\beta = 3.0$ respectively;
- Poisson yearly intensity $\lambda = 50$;
- Initial threshold value $L_0 = 2.0$.

We choose the threshold $L_0$ and scale parameter $\beta$ values so that after scaling by a common large factor they correspond to more or less typical values observed with real data. The standard deviations of the RWMH proposal transition kernel, adjusted to keep the RWMH acceptance probabilities close to 0.234, are $\sigma_\alpha = 0.2$, $\sigma_\beta = 0.3$, $\sigma_\lambda = 5$.

**Prior structure for MCMC.** The uniform prior distributions are used for all parameters, with the following bounds: $\alpha \in [0.1, 6]$, $\beta \in [0.1, 8]$ and $\lambda \in [0.1, 500]$. In the calculations below we checked that increasing the bounds does not lead to a material change in the estimates and thus the inferences are implied by data only.

### 5.2. MCMC and MLE results

For each of the chains obtained with MCMC, the first 1000 chain iterations are discarded, and all the results are obtained using the rest of the chain only. This is quite a common practice in MCMC modelling, used to increase the accuracy of results (as the chain needs to be run for a certain length before it starts to converge to the true values). Table 1 presents results for the posterior mean and standard deviation for each parameter using different chain lengths $K$. Figure 1 shows the histograms and scatter plots of the severity parameters $\alpha$, $\beta$ and intensity $\lambda$ obtained via the MCMC procedure for one of the data realisations over 5 years. Histograms were obtained from MCMC chains of length 5000 for each parameter. Note that the parameters are mutually dependent.

The standard deviations of the posterior distributions for the parameters reflect the parameter uncertainty due to the finite number of data points in the fitted sample while the finite number of iterations $K$ in the chain results in the numerical error in the MCMC estimates. The numerical standard error can be decreased by increasing $K$ and can be estimated as follows: split the chain into the non-overlapping bins; calculate the MCMC estimator using iterations for each bin; find the standard deviation of the obtained sample of estimators and divide by the square root of the number of bins. Table 1 shows the MCMC estimates and their numerical standard errors obtained using 100 bins. For our simulation parameters, the numerical standard errors are less then 2% when $K = 10^6$ which is small enough to be neglected. For all subsequent results, we use the chain length $K = 10^6$.

Table 2 compares ML estimates for marginal and joint estimations of the severity and frequency parameters. Maximization of the likelihood was performed using a numerical SPlus procedure based on a quasi-Newton method using the double dogleg step with the BFGS secant update to the Hessian.

We also show the results for the case of a mis-specified MLE that corresponds to the MLE fitting under a wrong assumption of constant threshold $L(t) \equiv L_0$. One observes from Table 3, that MCMC and ML estimators and corresponding standard deviations are in good agreement while mis-specified MLE leads to significantly different results.

As the data size increases, the parameter posterior mean converges to the true value and posterior standard deviation approaches zero. The same is observed for the MLE and its standard deviation. This convergence is suggested by the results in Table 4. One observes from Table 4, that MCMC and ML estimators are in good agreement. Note that standard deviations for MLEs are calculated under the Gaussian approximation for large data samples, so in general it is better to use MCMC.

The MCMC algorithm was implemented in C and computing time for $10^6$ iterations from a Markov chain in the case of a 5 year data sample was approximately 12 minutes on Intel Dual Xeon 2.4 GHz, 2 GB RAM.

**Full predictive distribution and quantile estimation (conditional on data)**
Using the simulation model specified above and applying the methodology described in Section 3.1, we estimate the full predictive distribution (conditional on data), see Figure 2, and calculate its $\hat{Q}_{0.999}^B$ quantile.

In Figure 2, the second histogram indicates the case of the data left truncated with the constant threshold (corresponding to the maximum threshold over the period of observations). The full predictive distribution (conditional on data) accounts for process uncertainty (that comes from the fact that we model losses using random process) and parameter uncertainty (that comes from the fact that the true parameter values are not known). We assume that numerical error due to finite number of simulations is negligibly small, as indicated by previous results.

The estimated value of the 0.999-quantile of the full predictive distribution is 1805.2, while the true 0.999-quantile of the aggregate loss distribution with the specified parameters is equal to 824.4 (the latter is calculated via FFT with advanced aliasing reduction techniques, the method described in Schaller and Temnov (2008)). The difference is due to parameter uncertainty for a finite sample size. Relying on the data left truncated with the constant threshold, the estimated value of the 0.999-quantile of the full predictive distribution is increased to 2367.0 because a smaller data set is used leading to a larger parameter uncertainty.

The impact of parameter uncertainty on the quantile can be also assessed by calculating the distribution (conditional on data) of $Q_{0.999}(\boldsymbol{\gamma})$, where $\boldsymbol{\gamma} = (\lambda, \boldsymbol{\beta})$ is from the posterior distribution $\pi(\boldsymbol{\gamma} | \mathbf{y})$, see Section 3.2. Using the sample $\boldsymbol{\gamma}^{(k)} = (\lambda^{(k)}, \boldsymbol{\beta}^{(k)})$ from the posterior distribution obtained via MCMC we calculate a sample of the quantile $Q_{0.999}(\boldsymbol{\gamma}^{(k)})$ via FFT and its histogram is presented on Figure 3. The empirical characteristics of the distribution for $Q_{0.999}(\boldsymbol{\gamma})$ are: mean = 1216.5; mode = 788.0; median = 1129.7; standard deviation = 764.4; 0.9-quantile = 3780; 0.95-quantile = 4697; 0.99-quantile = 5892. One can observe that the conditional distribution of the 0.999-quantile is considerably skewed, though its mean is less than the 0.999-quantile of the full predictive distribution.

The numerical error (due to the finite number of simulations $K$) of the quantile estimates can be calculated by forming a conservative confidence interval using the fact that the number of samples not exceeding the quantile has a Binomial distribution that can be approximated by the Normal distribution, see e.g. Shevchenko (2008). In the case of our simulation set up and $K = 10^6$, the numerical standard error is of the order of 1-2%.

**Unconditional distribution characteristics**

As the observation period increases the parameter posterior variance (the parameter uncertainty) should converge to zero and the quantile $\hat{Q}^B_{0.999}$ should converge to the true value 824.4, see Figure 4. For each selected observation period in Figure 4, 20 independent simulations were performed. The mean values of $\hat{Q}^B_{0.999}$ for each observation period are connected by solid line demonstrating the convergence.

## 6. Conclusions

In this paper, we considered modelling of a single operational risk in the case of data reported above a threshold varying in time. Of course, the assumption in this study is that missing losses and reported losses are realizations from the same distribution. Thus the approach should be used with caution especially if a large proportion of data is missing. In the latter case, it might be better to ignore missing data completely.

We have also demonstrated how the joint posterior distribution of the parameters can be estimated using MCMC and used to estimate the annual loss distribution accounting for both process and parameter uncertainties. In particular, we advocate the use of the *Random walk Metropolis-Hastings within Gibbs* algorithm which appeared to be efficient for the problem considered. Though, if the likelihood for the severity/frequency model can not be evaluated efficiently (as in the case of g-and-h distribution), then more advanced MCMC methods, such as Approximate Bayesian Computation should be used. Basically, combining the methodology proposed in the present paper with Approximate Bayesian Computation according to Peters and Sisson (2006), the generalization of the whole methodology to the case of distributions that do not allow an explicit likelihood is quite straightforward.

There are no conceptual problems to introduce time-varying threshold features into the model of many loss processes with dependence considered in Peters *et al* (2009b). However, while this multivariate model was presented for a general case involving frequencies and severities, the model calibration and required posterior densities were considered for the case of dependent frequencies only. In the case of threshold, the severity and frequency parameters should be estimated jointly. Deriving all required posteriors and implementation for this case are challenging tasks that will be addressed in future research.

In general, the following remarks can be made: estimation of the MLE uncertainties relies on asymptotic Gaussian approximation in the limit of large data samples; severity and frequency parameters should be estimated jointly; it is beneficial to use all available information in the fitting procedure.

In the present study we assumed constant non-informative priors so that inferences are implied by data only. It is important to note that informative priors can be used to incorporate external data and expert opinions into the model, see e.g. Lambrigger *et al* (2007). For a closely related credibility theory application in the context of operational risk, see Bühlmann *et al* (2007).

One of the features of the described Bayesian approach is that the variance of the posterior distribution $\pi(\boldsymbol{\gamma}|\mathbf{y})$ will converge to zero for a large number of observations. This means that the true value of the parameters will be known exactly. However, there are many factors (for example, political, economical, legal, etc.) changing in time that should not allow precise knowledge. One can model this by limiting the variance of the posterior distribution by some lower levels (e.g. 5%). This has been done in many solvency approaches for the insurance industry, see e.g. FOPI (2006) formulas (25)-(26). This can also be modelled by allowing the parameters be truly stochastic and possibly dependent as in Peters *et al* (2009b).

## Acknowledgements

The fist author thanks PRisMa Lab, Vienna University of Technology for financial travel assistance whilst completing aspects of this work at PRisMa Lab.
The authors thank the anonymous referee whose useful comments and remarks allowed improving the paper.

| K | Posterior mean (posterior stdev) | | | Numerical standard error | | |
|---|---|---|---|---|---|---|
| | $\alpha$ | $\beta$ | $\lambda$ | $\alpha$ | $\beta$ | $\lambda$ |
| $10^4$ | 2.18 (0.61) | 3.25 (0.79) | 60.6 (8.8) | 0.06 | 0.07 | 1.8 |
| $10^5$ | 2.15 (0.68) | 3.22 (0.84) | 58.9 (9.65) | 0.04 | 0.05 | 1.0 |
| $10^6$ | 2.12 (0.73) | 3.18 (0.91) | 56.5 (10.4) | 0.018 | 0.02 | 0.4 |

**Table 1.** MCMC results for different chain length $K$ in the case of data simulated over 5 years from $Poisson(\lambda = 50)$ - $Pareto(2,3)$ and truncated below $L(t) = L_0 \exp(rm)$, $m-1 \le t < m$, $m = 1,...,5$.

| $\lambda$ | marginal MLE (stdev) | | | joint MLE (stdev) | | | MCMC mean (stdev) | | |
|---|---|---|---|---|---|---|---|---|---|
| | $\alpha$ | $\beta$ | $\lambda$ | $\alpha$ | $\beta$ | $\lambda$ | $\alpha$ | $\beta$ | $\lambda$ |
| 1 | 1.16 (2.50) | 2.18 (3.40) | 0.78 (0.94) | 1.29 (2.38) | 2.17 (3.12) | 0.71 (0.90) | 1.25 (2.15) | 2.16 (3.4) | 0.73 (0.88) |
| 10 | 2.25 (1.33) | 3.61 (1.87) | 16.5 (8.4) | 2.23 (1.25) | 3.53 (1.76) | 16.1 (7.8) | 2.20 (1.25) | 3.47 (1.55) | 14.4 (7.9) |
| 100 | 1.88 (0.85) | 2.7416 (1.04) | 107.05 (20.5) | 1.89 (0.85) | 2.74 (1.04) | 106.2 (19.9) | 1.91 (0.85) | 2.73 (1.03) | 106.5 (19.7) |
| 500 | 2.09 (0.46) | 3.2636 (0.60) | 487.1 (75.3) | 2.09 (0.45) | 3.26 (0.59) | 485.8 (75.0) | 2.12 (0.43) | 3.15 (0.59) | 491 (74.2) |
| 1000 | 2.06 (0.33) | 3.1255 (0.46) | 1016 (40.4) | 2.06 (0.33) | 3.13 (0.46) | 1014.5 (40.3) | 2.05 (0.33) | 3.07 (0.46) | 1009 (38.8) |

**Table 2**. The MLEs for the Pareto severity parameters $(\alpha, \beta)$ using joint and marginal estimations. The data were sampled over 5 years from $Poisson(\lambda)$ - $Pareto(2,3)$ and truncated below $L(t) = L_0 \exp(rm)$, $m-1 \le t < m$, $m = 1,...,5$.

|   | ML | Mis-specified ML | Bayesian posterior |
|---|---|---|---|
|   | MLE (stdev) | MLE (stdev) | mean (stdev) |
| $\alpha$ | 2.23(0.75) | 3.38(0.91) | 2.12(0.73) |
| $\beta$ | 3.14(0.90) | 3.87(1.04) | 3.18(0.91) |
| $\lambda$ | 55.0(12.4) | 76.5(17.5) | 56.5(10.4) |

**Table 3.** MLE and Bayesian estimators in the case of data simulated over 5 years from the $Poisson(50)$-$Pareto(2,3)$ and truncated below $L(t)=L_0 \exp(rm)$, $m-1 \le t < m$, $m=1,...,5$. Mis-specified ML are MLE obtained under assumption of constant threshold $L(t) \equiv L_0$.

| M | J | posterior mean (stdev) | | | MLE (MLE stdev) | | |
|---|---|---|---|---|---|---|---|
|   |   | $\alpha$ | $\beta$ | $\lambda$ | $\alpha$ | $\beta$ | $\lambda$ |
| 1 | 43 | 2.33(0.97) | 3.25(1.15) | 61.0(24.0) | 2.42(0.95) | 3.38(1.25) | 62.2(28.1) |
| 2 | 85 | 1.72(0.84) | 2.78(1.02) | 37.0(15.5) | 1.66(0.83) | 2.75(1.08) | 58.6(21.5) |
| 5 | 196 | 2.12(0.73) | 3.18(0.91) | 56.5(10.4) | 2.23(0.75) | 3.14(0.90) | 55(12.4) |
| 10 | 311 | 1.89(0.59) | 2.84(0.84) | 44.5(8.2) | 1.87(0.61) | 2.82(0.85) | 50.5(9.7) |
| 12 | 389 | 1.93(0.52) | 2.82(0.79) | 45.2(7.0) | 1.9(0.63) | 2.76(0.78) | 42(8.5) |
| 14 | 430 | 2.15(0.48) | 3.08(0.74) | 53.5(6.2) | 2.09(0.58) | 3.17(0.73) | 49(8.9) |
| 16 | 485 | 1.95(0.45) | 2.89(0.69) | 48.7(4.7) | 1.94(0.51) | 2.94(0.67) | 47.5(7.7) |
| 18 | 521 | 1.91(0.42) | 2.96(0.65) | 48.4(4.1) | 1.95(0.46) | 2.92(0.65) | 52.5(6.9) |
| 20 | 545 | 2.04(0.38) | 3.05(0.52) | 51.4(3.8) | 2.07(0.42) | 3.11(0.58) | 49.2(5.4) |

**Table 4.** MCMC and MLE results as the data sample size increases. $M$ is the number of simulated years, $J$ is the total number of events occurred during $M$ years. The data are simulated from $Poisson(50)$-$Pareto(2,3)$.

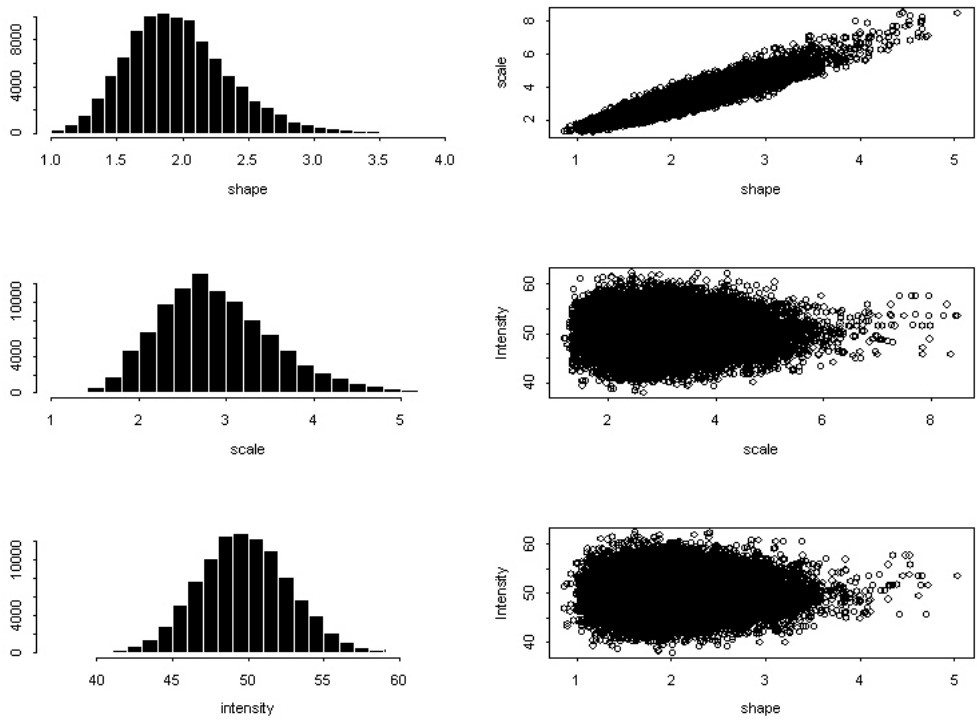

**Figure 1**. Histograms and scatter plots of the MCMC samples of the model parameters (Pareto shape, Pareto scale and Poisson intensity), in the case of fie year data sample.

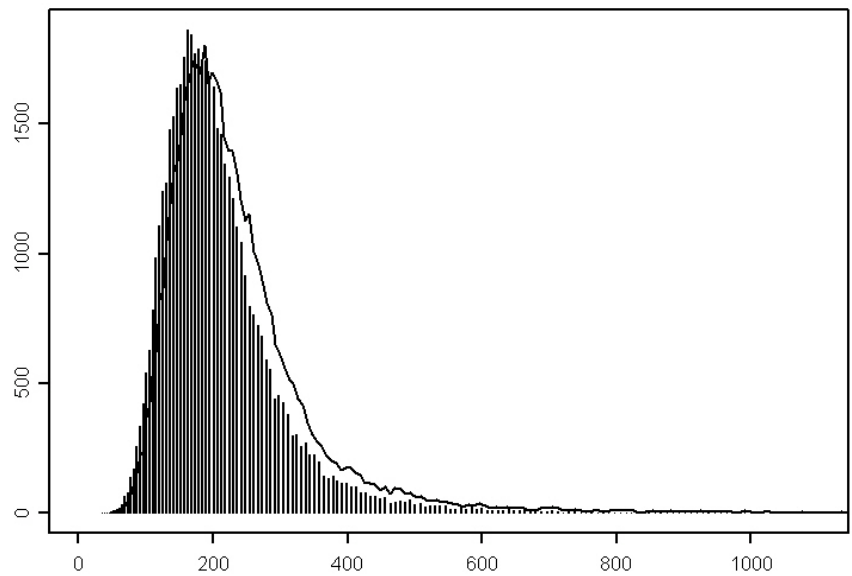

**Figure 2**. Histograms of the full predictive distribution $h(z|\mathbf{y}) = \int h(z|\boldsymbol{\gamma})\pi(\boldsymbol{\gamma}|\mathbf{y})d\boldsymbol{\gamma}$ (conditional on data $\mathbf{y}$) using MCMC of length 50,000. Histogram with black ticks was obtained using the whole data sample. The second histogram (outlined by a solid line) corresponds to the reduced sample of data truncated above the maximum threshold over the period of observations.

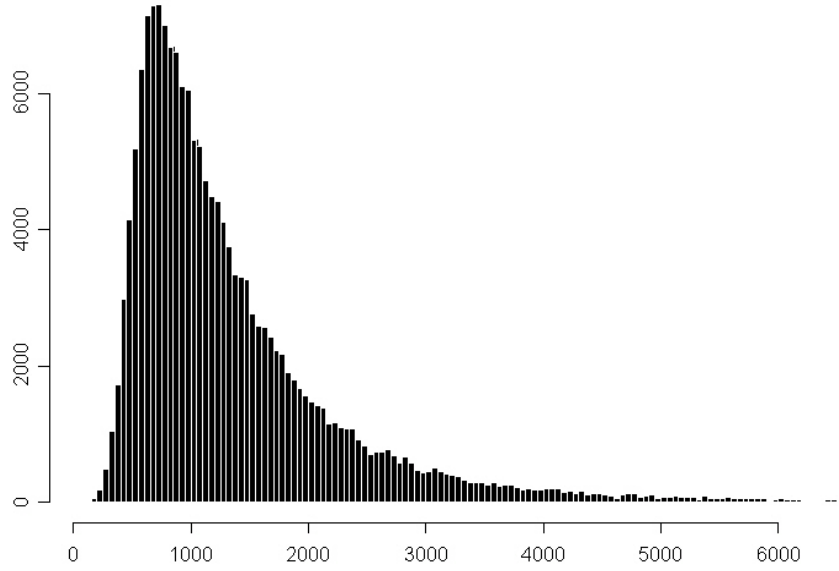

**Figure 3**. Histogram of the 0.999 quantile $Q_{0.999}(\boldsymbol{\gamma}^{(k)})$, $k=1,\ldots,10^4$, where $\boldsymbol{\gamma}^{(k)}$ are samples from the posterior $\pi(\boldsymbol{\gamma}\mid\mathbf{y})$.

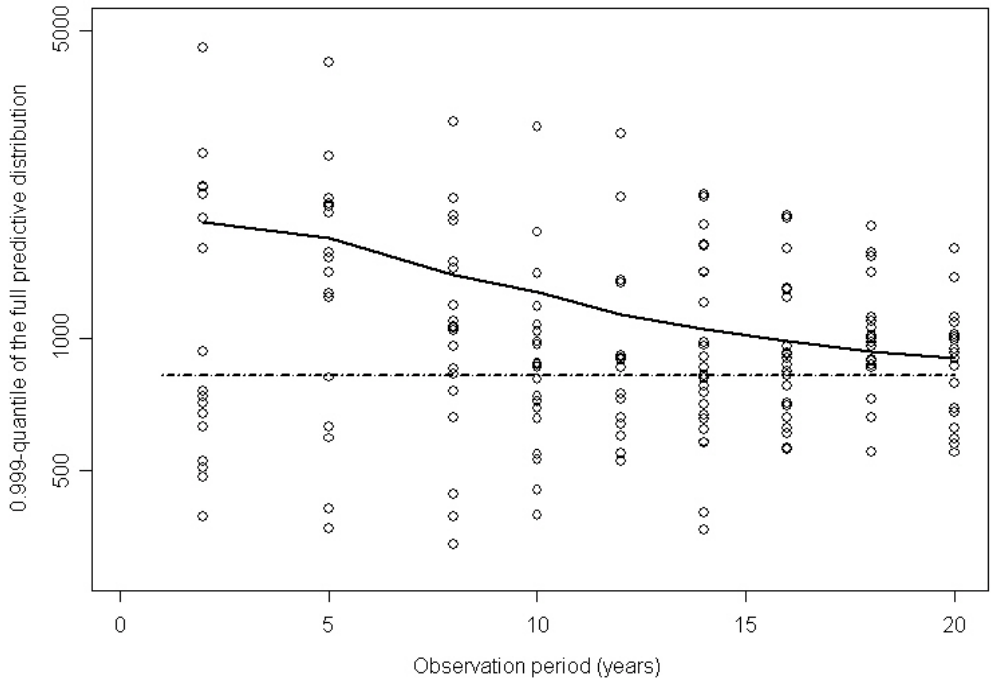

**Figure 4**. The 0.999 quantile, $\hat{Q}_{0.999}^{B}$, of the full predictive distribution vs observation period for 20 independent data realizations. Averages over realizations for each observation period are connected by thick solid line. Dotted line indicates the 0.999 quantile of the aggregate loss distribution with the true parameters.